\pdfoutput=1 
\documentclass{JINST}
\usepackage{multirow}
\usepackage{upgreek}

\title{Time resolution below 100\,ps for the SciTil detector of PANDA employing SiPM}

\author{S. E. Brunner$^a$\thanks{Corresponding author.},
L. Gruber$^a$~,
J. Marton$^a$~,
H. Orth$^b$~
and K. Suzuki$^a$\\
\llap{$^a$}Stefan Meyer Insitute for Subatomic Physics, Austrian Academy of Sciences,\\
  Boltzmanngasse 3, 1090 Vienna, Austria\\
\llap{$^b$}GSI Helmholtzinstitut f\"ur Schwerionenforschung,\\
  Planckstra\ss{}e 1, 64291 Darmstadt, Germany\\
E-mail: \email{stefan.enrico.brunner@oeaw.ac.at}}

\abstract{The barrel time-of-flight (TOF) detector for the $\overline{\mathrm{P}}$ANDA experiment at FAIR in Darmstadt is planned as a scintillator tile hodoscope (SciTil) using 8000 small scintillator tiles. It will provide fast event timing for a software trigger in the otherwise trigger-less data acquisition scheme of $\overline{\mathrm{P}}$ANDA, relative timing in a multiple track event topology as well as additional particle identification in the low momentum region. The goal is to achieve a time resolution of $\sigma\simeq$\,100\,ps. We have conducted measurements using organic scintillators coupled to Silicon Photomultipliers (SiPM). The results are encouraging such that we are confident to reach the required system time resolution.}

\keywords{Scintillation detector; timing detector; fast timing; time resolution; Silicon Photomultiplier (SiPM); PANDA}

\begin{document}

\section{Introduction}\label{sec:intro}
The Scintillator Tile Hodoscope (SciTil) is a proposed sub-detector of the planned $\overline{\mathrm{P}}$ANDA experiment~\cite{panda} situated at FAIR~\cite{fair}. The tasks of the SciTil detector are particle identification for slow particles (below 700 MeV/c) in combination with a central tracker~\cite{gillitzer2011}, relative time-of-flight information, event timing, conversion detection and charge discrimination in front of the electromagnetic calorimeter. Furthermore, it would increase the detection probability for $\overline{\Xi}$ by monitoring slow Kaon decay products and help to deconvolute particle tracks. The space for the barrel SciTil is limited to 2\,cm in radial direction including the support structure. Requirements for the detector are minimum use of material and a time resolution of $\sigma\simeq$\,100\,ps.

A basic layout of the detector has been proposed in Ref.~\cite{scitil}. It suggests tiles made out of small organic scintillators with sizes of $\sim30\times30\times5\,\mathrm{mm}^3$, attached to Silicon Photomultipliers (SiPM) with a sensitive area of 3$\times$3\,mm$^2$, see Fig.~\ref{fig:dsipm_setup}, left hand side. The main reasons for choosing organic scintillators are their fast response (short rise- and decay-times) and their high light yield. SiPM provide advantageous properties such as good timing, compactness, high photon detection efficiency (PDE) and operation in magnetic fields which will be 2\,T in the $\overline{\mathrm{P}}$ANDA target spectrometer.

In order to achieve the required time resolution of $\sigma\simeq$\,100\,ps, research on the detector design was divided into two major parts, the scintillator and the photodetector. The first part includes the choice of the optimally suited scintillator material, the size and shape of the scintillator, as well as finding the optimal position to attach the photodetectors onto the scintillator. The second part contains the identification of the SiPM with the best time resolution among a variety of manufacturers and the determination of optimal operating conditions for the expected photon pulse shapes emitted by the scintillators. 

\section{Scintillator time resolution}\label{sec:scint}
\subsection{Method} There are several parts contributing to the time resolution of a scintillation detector, e.g. the scintillator type, the photodetector and the electronics. Regarding the scintillator itself, the material properties, e.g. rise- and decay-times and light output, are directly influencing the time resolution. The coupling between scintillator and photodetector is important such that no photons are lost at the transition. Furthermore, the matching between the emission spectrum of the scintillator and the spectral sensitivity of the SiPM is a critical point.

\begin{figure}[tbp]
  \centering
  \includegraphics[width=.33\textwidth]{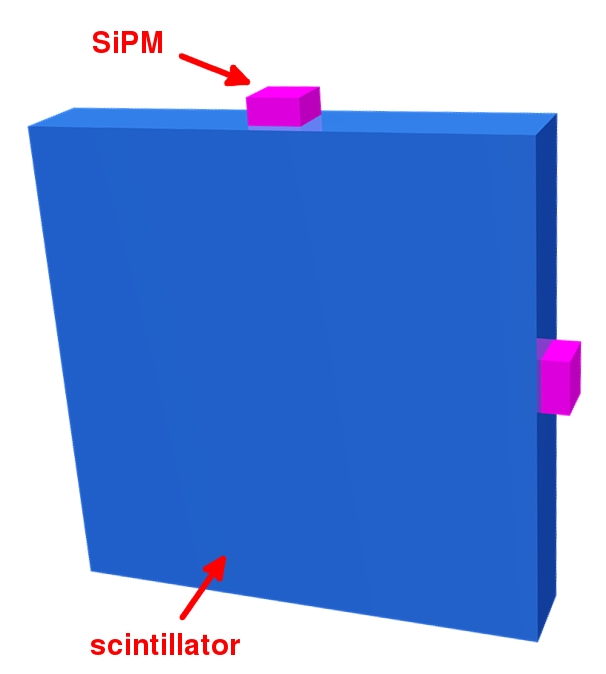}
  \includegraphics[width=.49\textwidth]{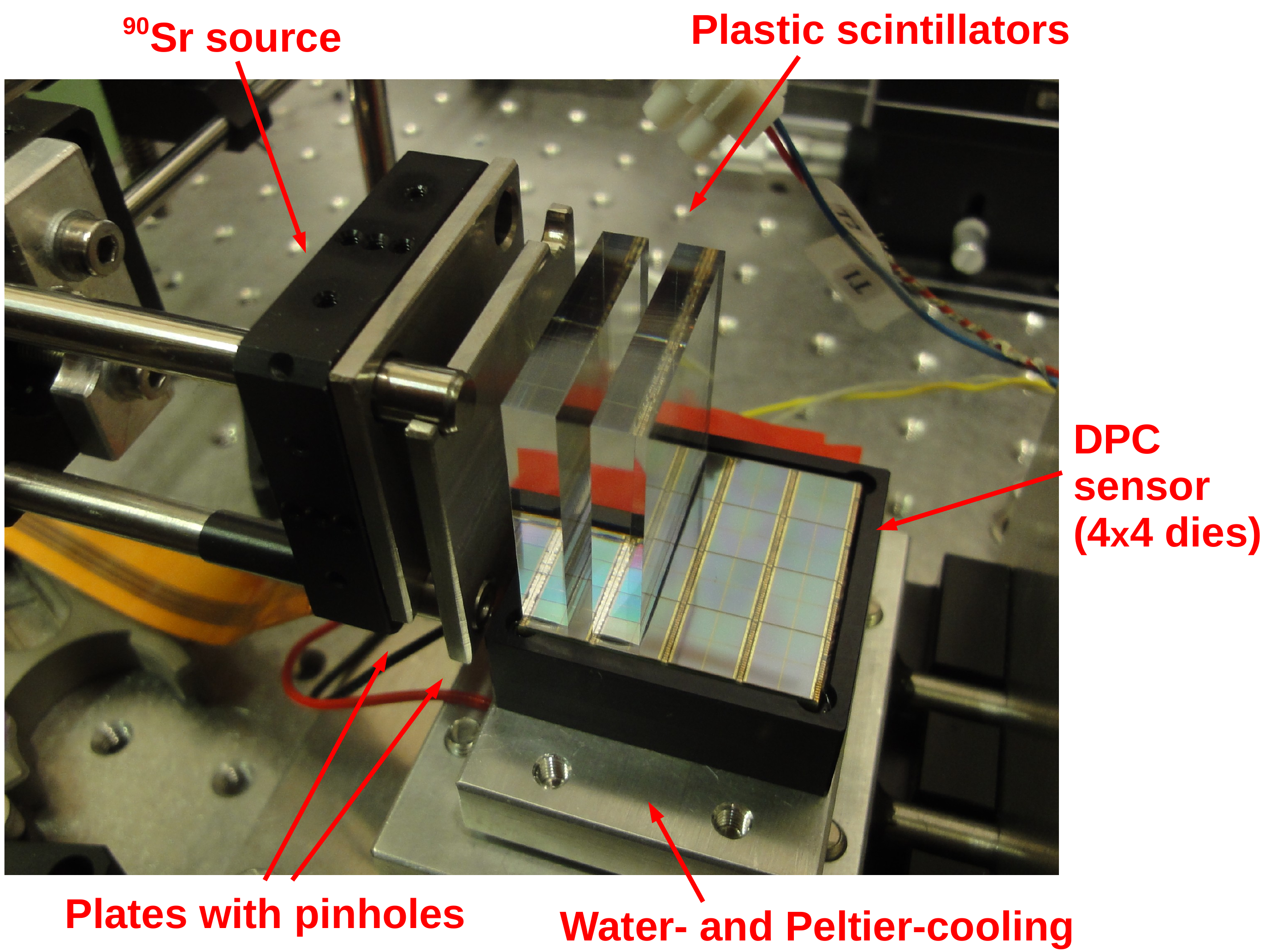}
  \caption{Left: conceptual design for the layout of one scintillation tile for SciTil with a size of \mbox{$\sim30\times30\times5\,\mathrm{mm}^3$}. Right: setup used to evaluate the time resolution of plastic scintillator tiles. The scintillators are read out using the Philips DPC, which consists of 16 individual die sensors arranged in a 4$\times$4 matrix.}
  \label{fig:dsipm_setup}
\end{figure}

The geometry of the scintillator influences the time resolution in a way that light path variations from the point of photon creation to the photodetector add time spread to the signal. For SciTil a square tile size of 20\,mm to 30\,mm is proposed~\cite{scitil}, in order to achieve a time resolution of about 100\,ps on the one hand and to keep the number of channels low on the other hand. A thickness in the order of 5\,mm is sufficient to create $\sim$10\,k photons in the proposed plastic scintillators for minimum ionizing particles (MIP). Simulations have shown that a time resolution below 100\,ps is feasible when detecting more than 100 photons~\cite{scitil}. This can be reached when attaching two SiPM, with a sensitive area of 3\,x\,3\,mm$^2$ each, to the rim of a tile. To increase the number of detected photons and improve time resolution, the ideal position of the photodetector on the tile has to be found.

In order to study the scintillator time resolution and the individual parameters described above, a simple setup was used, see Fig.~\ref{fig:dsipm_setup}, on the right hand side. A $^{90}$Sr source provides electrons up to an energy of 2.28\,MeV. The electrons were collimated using two steel plates with 3\,mm thickness and centered holes of 2\,mm diameter. These pinholes and the source could be moved with \mbox{$\upmu$m-resolution} in two dimensions in order to define the beam position. As photodetector, the Digital Photon Counter (DPC) from Philips~\cite{dsipm} was used. Being a fast photodetector with large active area and position sensitivity, the DPC is suited for such comparative studies and for testing position dependency of the time resolution by simply placing the plastic scintillator tiles on top of the DPC using optical grease in between. Like the analog device, the DPC consists of an array of Single Avalanche Photo Diodes (SPADs). In contrast to the conventional SiPM, where the output signal 
corresponds to the analog sum of individual SPAD pulses, the DPC output is the digital sum of trigger bins with additional digital time stamps from the TDCs. The DPC has an active area of 32.6\,$\times$\,32.6\,mm$^2$ and consists of 16 independent die sensors, arranged in a 4$\times$4 matrix (see Fig.~\ref{fig:dsipm_setup} on the right hand side). A die is sub-divided into a 2$\times$2 pixel matrix. One pixel has a sensitive area of 3.2\,x\,3.9\,mm$^2$, which comes close to the size of a conventional SiPM. Depending on the sensor type (DPC-6400 or DPC-3200), each pixel consists of 6400 or 3200 SPADs, respectively. The user can define how many dies, pixels or individual SPADs to activate for data acquisition. After occurrence of an event, the number of breakdowns (photon count) and a single time stamp per die corresponding to the trigger time is saved.

The DPC was operated at 3\,V above the breakdown voltage for all measurements and the trigger level was set to 1 photon, in order to use the time stamp of the first detected photon. In case of analog SiPM the optimum threshold would be higher due to statistical effects \cite{Gundacker2013}. The setup was mounted inside a dark box and the temperature was stabilized at 20\,$^\circ$C using water- and Peltier-cooling.  In this experimental study we used the following plastic scintillators: 1 piece of BC-408 with size of $30\times30\times4\,\mathrm{mm}^3$, 2 pieces of BC-408 with size of $25\times25\times5\,\mathrm{mm}^3$, both from Saint-Gobain Crystals and 2 pieces of EJ-228 ($30\times30\times5\,\mathrm{mm}^3$) from Eljen Technology. EJ-228 has similar physical properties as BC-418. The main parameters of the scintillators are summarized in Table~\ref{tab:scintpara}.

\begin{table}[tbp]
  \caption{Scintillation properties taken from data sheets~\cite{bicron, eljen}.}
  \label{tab:scintpara}
  \smallskip
  \centering
  \begin{tabular}{|lcc|}
    \hline
    Scintillator&BC-408&EJ-228\\
    Light yield [photons/MeV]&10,000&10,200\\
    Rise time [ns]&0.9&0.5\\
    Decay time [ns]&2.1&1.4\\
    Pulse width (FWHM) [ns]&2.5&1.2\\
    Wavelength of max. emission [nm]&425&391\\
    \hline
  \end{tabular}
\end{table}

\subsection{Results}\label{sec:scint_results}
\begin{figure}[tbp]
  \centering
  \includegraphics[width=0.50\textwidth]{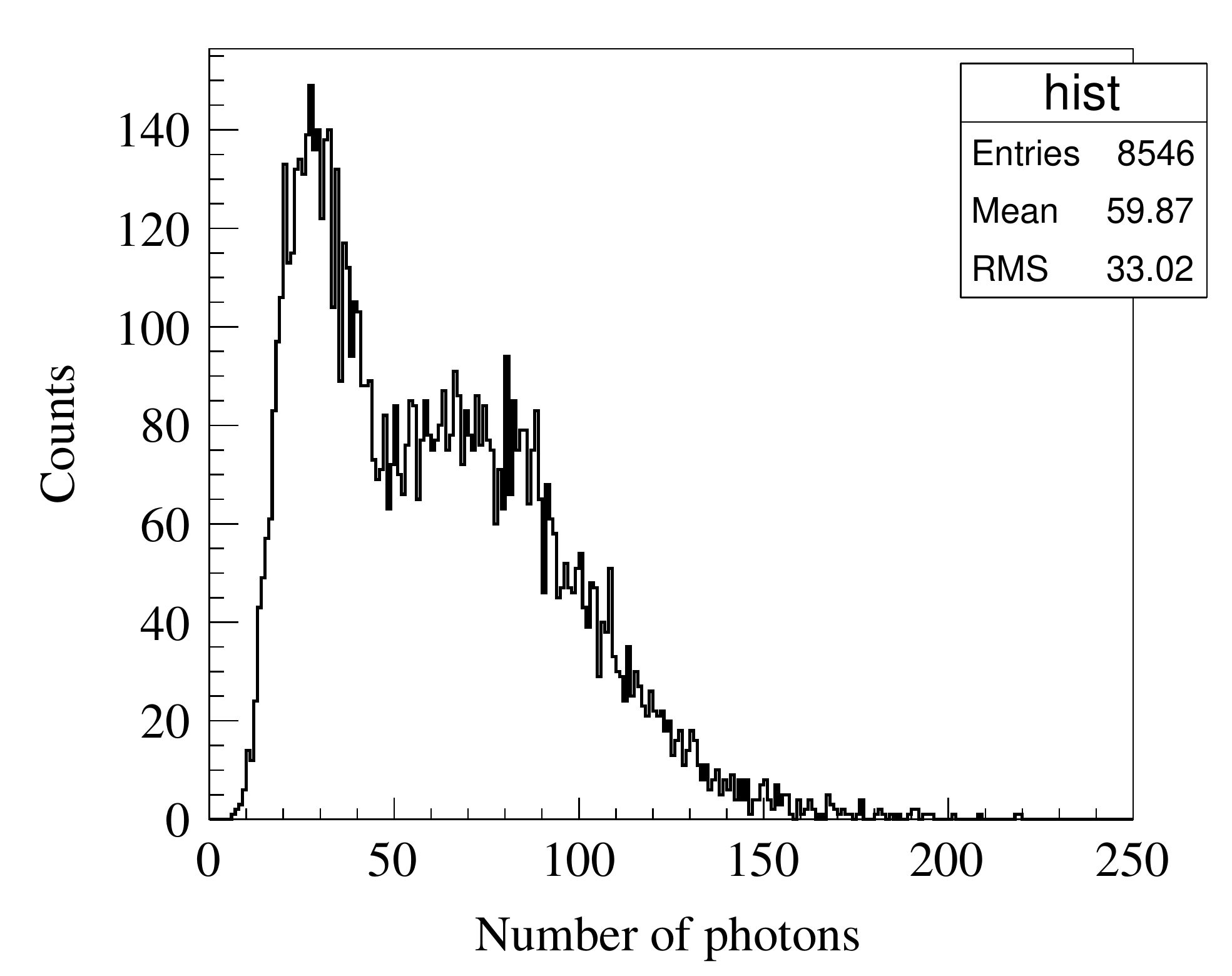}
  \caption{$^{90}$Sr spectrum measured with a BC-408 scintillator ($30\times30\times4\,\mathrm{mm}^3$) and a DPC-6400. The histogram shows the photon number detected on one die (2 pixels activated).}
  \label{fig:Sr90spectrum}
\end{figure}

\begin{figure}[tbp]
  \centering
  \includegraphics[width=.43\textwidth]{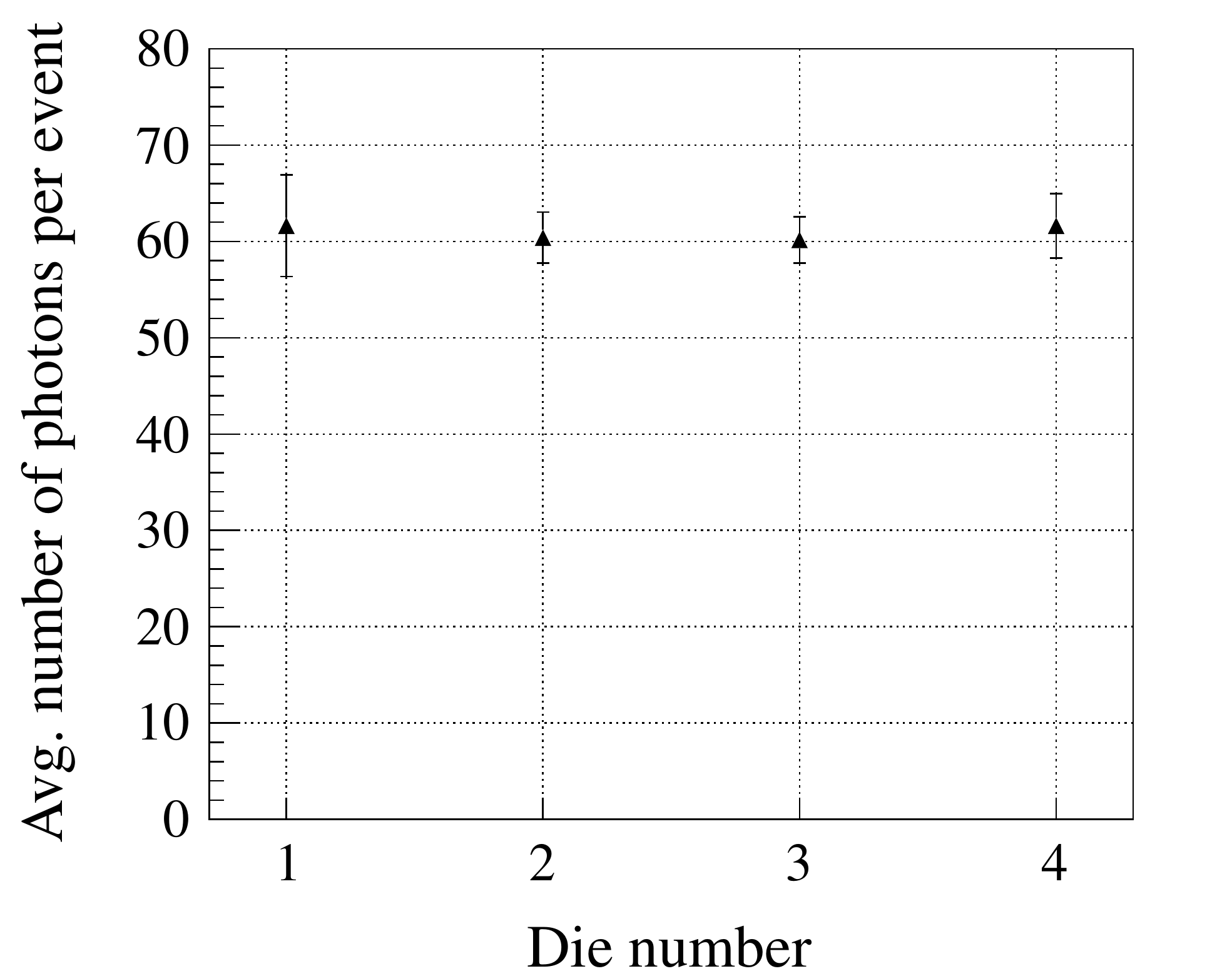}
  \includegraphics[width=.43\textwidth]{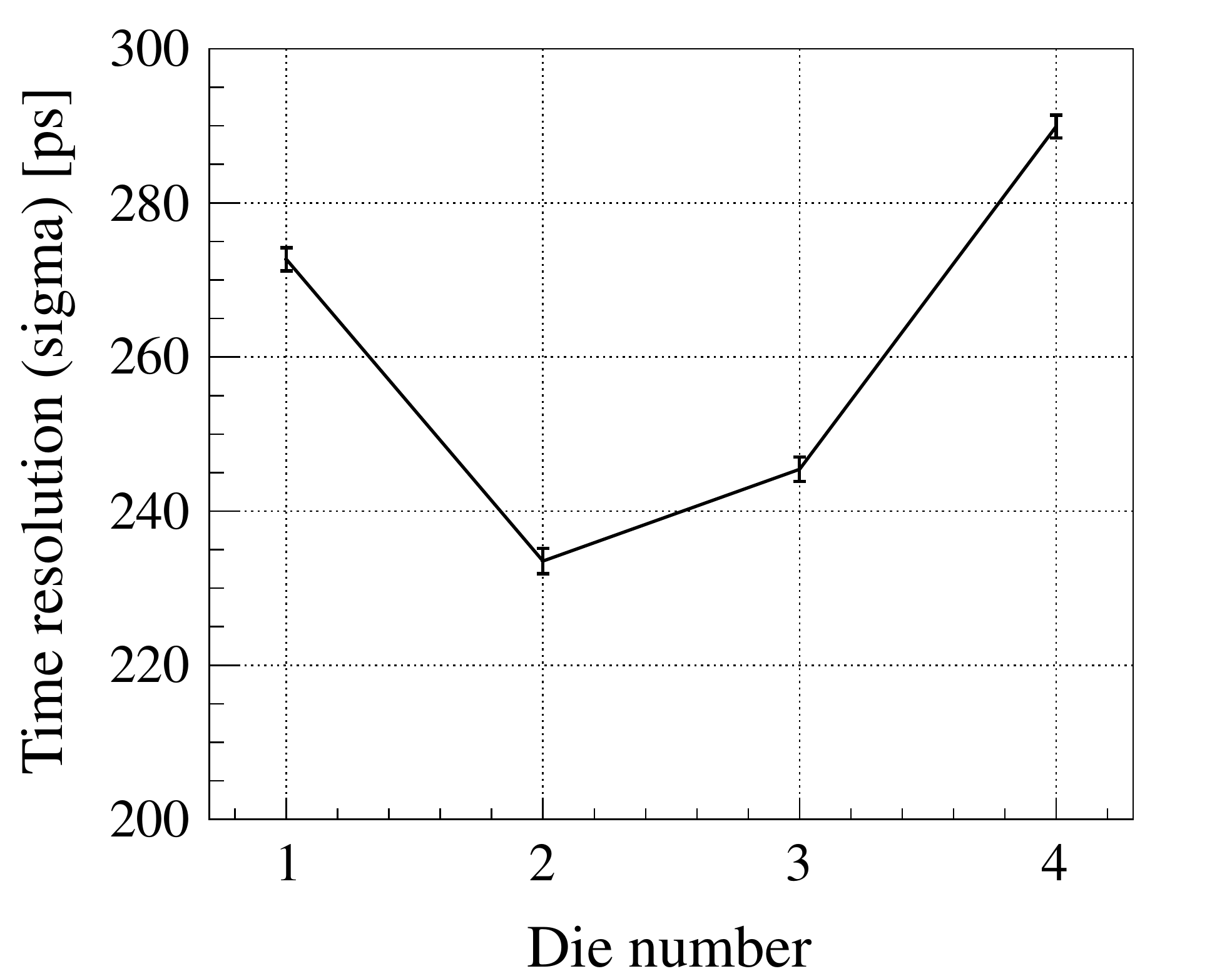}
  \caption{The plot on the left shows the photon distribution measured at the rim of the scintillator (BC-408, $30\times30\times4\,\mathrm{mm}^3$). Die number 1 and 4 correspond to the dies located at the edge of the scintillator tile, die number 2 and 3 to the central dies. On the right, one can see the corresponding time resolution. The values are mean numbers of 25 different beam positions.}
  \label{fig:photdistribution}
\end{figure}

First we measured the number of detected photons and evaluated the photon distribution along the rim of the scintillator. Therefore, a BC-408 scintillator tile ($30\times30\times4\,\mathrm{mm}^3$) was coupled to the DPC-6400 and read out by 4 dies. Since only half of a die's active area was covered by the thin scintillator, only 2 out of 4 pixels per die were activated during data acquisition.  The measurement was performed by moving the $^{90}$Sr-source position two-dimensionally in steps of 5\,mm across the entire scintillator surface, counting the number of detected photons at each position. Fig.~\ref{fig:Sr90spectrum} shows the photon number spectrum for electrons directed onto the center of the square scintillator surface for two pixels activated. The photon counts of individual die sensors are shown in Fig.~\ref{fig:photdistribution} (left hand side). The plot shows a mean photon count of all 25 positions. Die number 1 and 4 are located at the edge of the scintillator tile, die number 2 and 3 at the 
center. Evidently, the photons are rather equally shared among the different dies, demonstrating scintillation light to be equally distributed over the rim of the scintillator. Since each die gives a time stamp at the instant of trigger generation (detection of the first arriving photon), one can disentangle the time resolution of each die using all possible die combinations. On the right hand side of Fig.~\ref{fig:photdistribution} one can see that the time resolution improves for dies positioned in the center. Again, the values are mean numbers for 25 source positions. The asymmetry of the graph can be explained by an instrumental asymmetry, caused by the positioning of the scintillator on the active area of the DPC. In the analysis, all events were considered without energy cuts on the photon spectra. The results show that there is a position dependence of the time resolution, indicating that the best timing can be achieved by placing the photodetector at the center of the scintillator rim. For equally 
distributed source positions, the light path variations inside the scintillator from the point of photon creation to the photodetector are on average smaller when the detector is placed in the center and thus, the arrival time jitter of the detected photons is smaller.        

\begin{figure}[tbp]
  \centering
  \includegraphics[width=.43\textwidth]{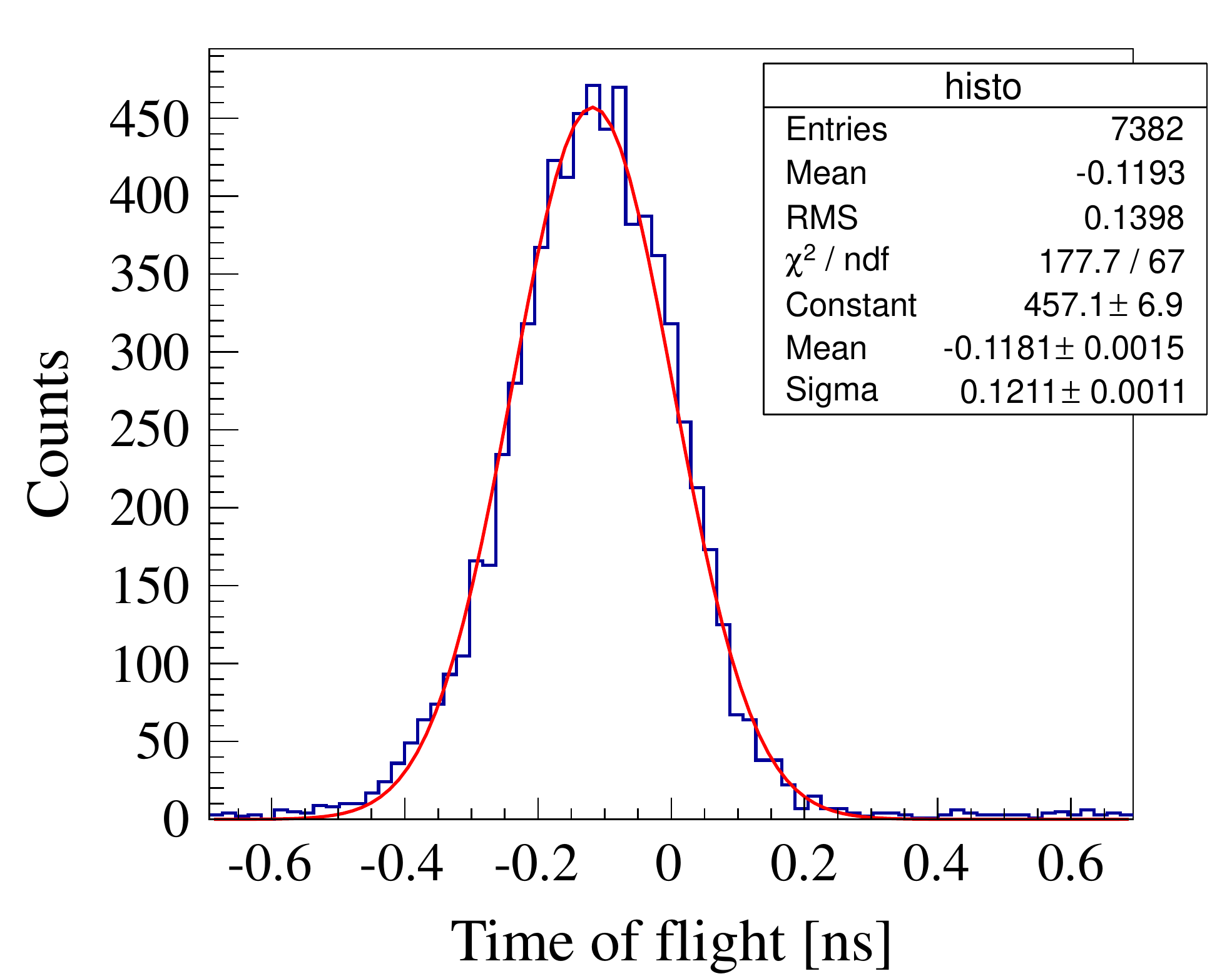}
  \includegraphics[width=.43\textwidth]{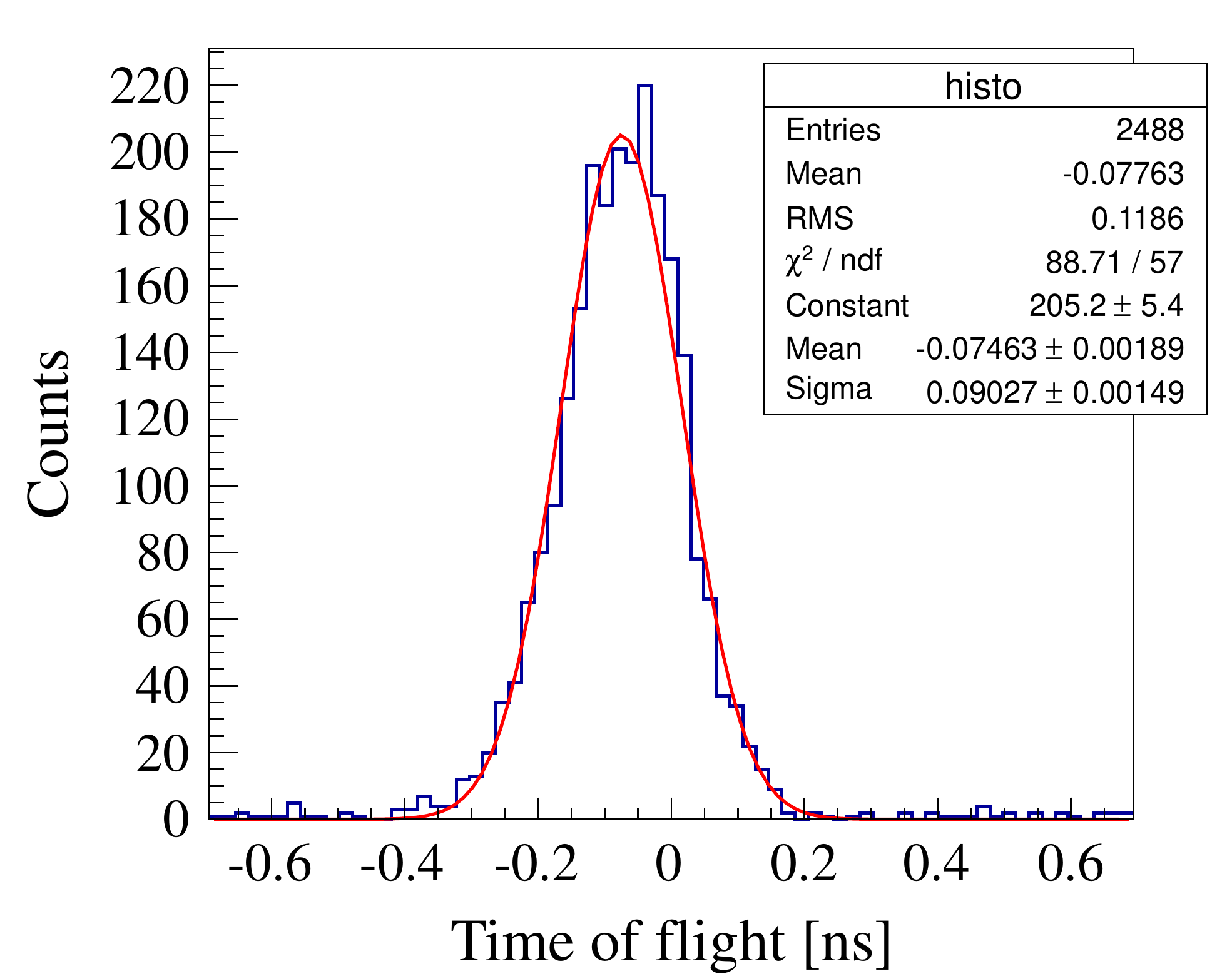}
  \caption{Time-of-flight spectrum between two tiles of BC-408 with size of $25\times25\times5\,\mathrm{mm}^3$ (left) or EJ-228 with size of $30\times30\times5\,\mathrm{mm}^3$ (right). The spectra were fit using a Gaussian distribution.}
  \label{fig:CTR}
\end{figure}

In order to increase the detected number of photons, we exchanged the DPC model and used the DPC-3200 for further measurements. This model consists of 3200 SPADs per pixel and provides a higher PDE compared to the DPC-6400 due to a higher fill factor. Two plastic scintillator tiles were placed on the sensitive area of the DPC, covering two rows of dies (see Fig.~\ref{fig:dsipm_setup}). In each row only the central two dies (two pixels each) were activated and used for read out. The source position was directed onto the center of the scintillator surface. By putting the two scintillator tiles in coincidence (in total 4 dies), only high energy events ($\Delta$E > 0.8\,MeV in the first tile) were selected. For each die, the arrival time of the first photon was saved. The start and stop signals were created by taking the mean of the two time stamps for the first and second tile, respectively. Fig.~\ref{fig:CTR} shows the time-of-flight spectra using two tiles of BC-408 with a size of $25\times25\times5\,\mathrm{
mm}^3$ and EJ-228 with a size of $30\times30\times5\,\mathrm{mm}^3$, respectively. With the latter, a time-of-flight resolution of 90\,ps (sigma) was achieved, much better than with the BC-408 tile. The EJ-228 scintillator has a larger surface. Hence, travel path variations inside the scintillator are larger and the number of detected photons is a factor 1.3 to 1.4 smaller at comparable light yield, compared to the BC-408 scintillator. However, the timing properties are superior because of shorter rise- and decay-times (see Table~\ref{tab:scintpara}).         

\begin{figure}[tbp]
  \centering
  \includegraphics[width=0.43\textwidth]{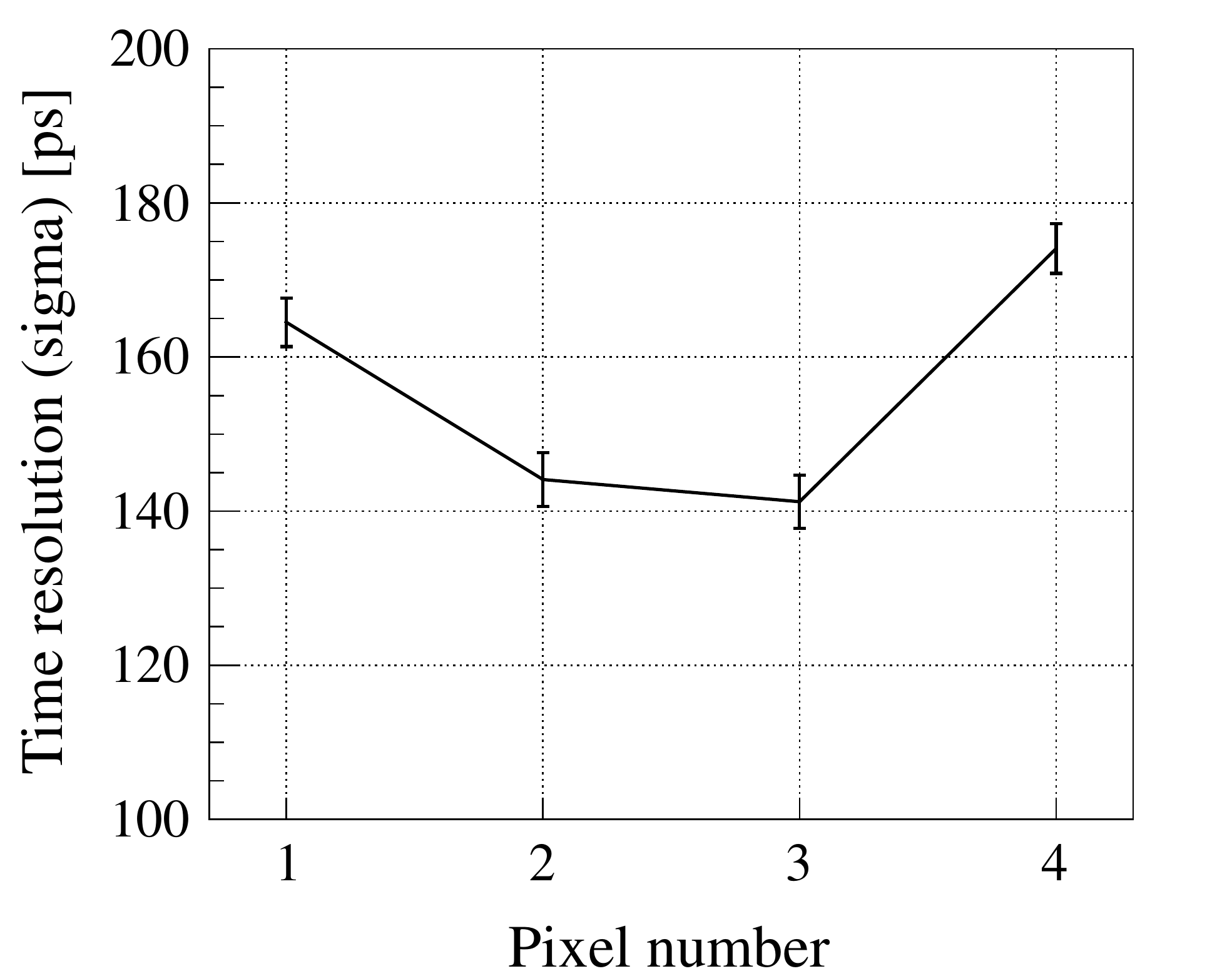}
  \includegraphics[width=0.43\textwidth]{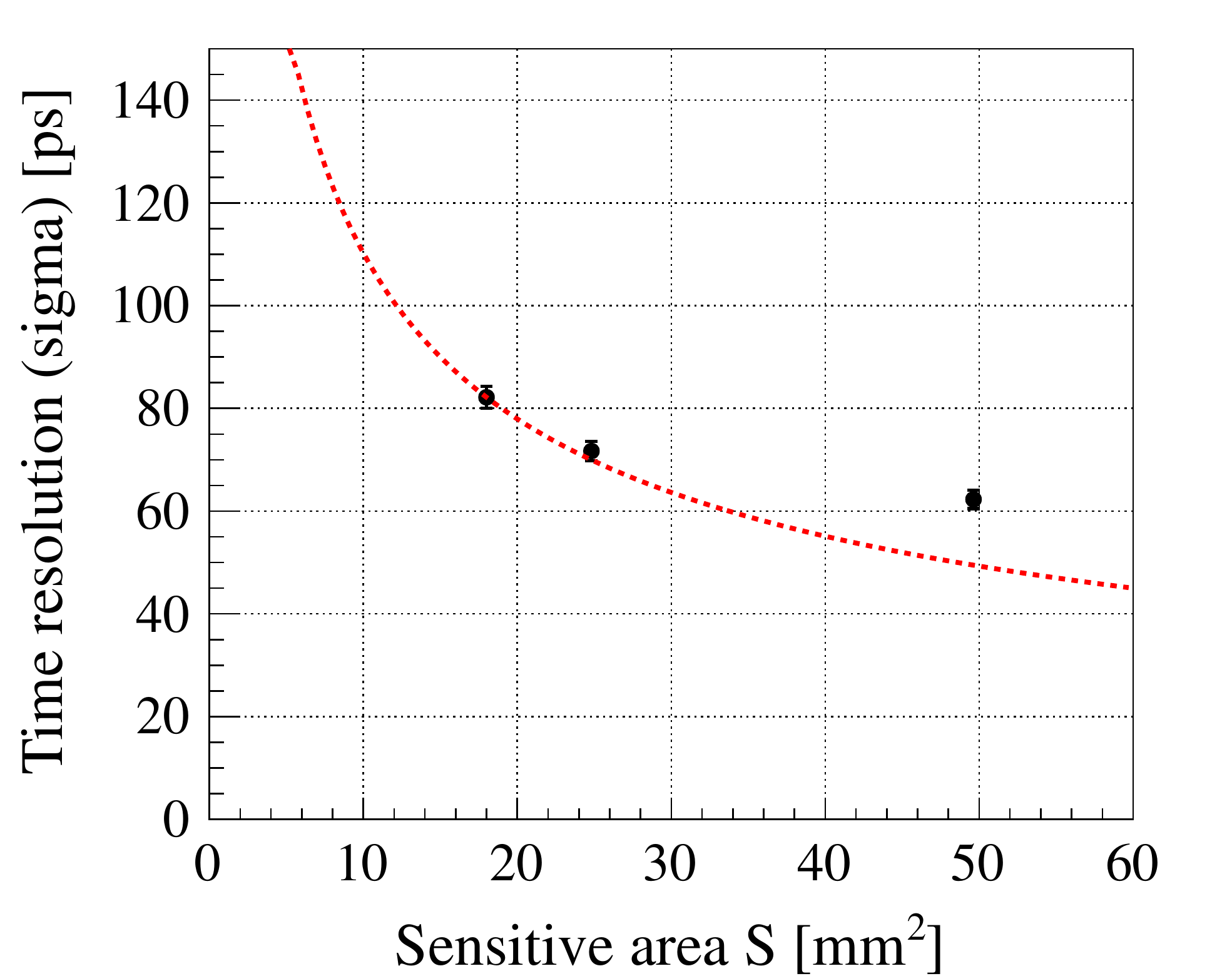}
  \caption{Time resolution of BC-408 ($25\times25\times5\,\mathrm{mm}^3$) read out with single pixels (left). Pixel number 1 and 4 correspond to pixels located at the edge of the scintillator tile, pixel number 2 and 3 to pixels in the center. On the right hand side, the time resolution of one scintillator tile (EJ-228, $30\times30\times5\,\mathrm{mm}^3$) as a function of the sensitive sensor surface (S) is plotted. The black dots are the data points, the dashed curve indicates the time resolution expected when scaling with 1/$\sqrt{S}$, normalized to the point S\,=\,18\,mm$^2$. The deviation at large sensitive area could be explained by the skew of the trigger network scheme of the DPC.}
  \label{fig:pixeltiming}
\end{figure}

Since it is planned to finally read out the scintillator tiles with conventional SiPM, we decreased the detection area on the DPC by switching on only individual pixels, which have similar sensitive surface compared to analog sensors. One pixel of the DPC has an active area of about 12\,mm$^2$. The measurement was performed using 2 tiles of BC-408 with a size of $25\times25\times5\,\mathrm{mm}^3$. The first scintillator tile is again read out using two dies (two pixels per die), for the second tile only single pixels are used. Fig.~\ref{fig:pixeltiming} (left) shows the time resolution of the second scintillator tile, read out by individual pixels. Pixel number 1 and 4 correspond to pixels located at the edge of the scintillator tile, pixel number 2 and 3 to pixels in the center. Using both center pixels for read out instead of one, the time resolution was improved to 115\,ps (sigma).

The sensitive area of the DPC was further decreased to 3$\times$3\,mm$^2$ by switching off individual SPADs. On the right hand side of Fig.~\ref{fig:pixeltiming}, one can see the time resolution of the EJ-228 tile in dependence of the activated sensitive area used for read out. The time resolution improves with increasing sensitive area S, since the number of detected photons $\mathrm{N_{ph}}$ is dependent on the sensitive area. From statistics, one could expect that the time resolution improves with 1/$\sqrt\mathrm{N_{ph}}$, as indicated by the dashed line in Fig.~\ref{fig:pixeltiming} (right). For increasing sensitive area the skew of the trigger network scheme seems to become a prominent factor influencing the time resolution~\cite{Frach2010}. This is visible for the deviation of the data point from the dashed curve at S\,=\,50\,mm$^2$.

\section{SiPM time resolution}\label{sec:sipm}
\subsection{Method} 
\begin{figure}[tbp]
  \centering
  \includegraphics[width=.38\textwidth]{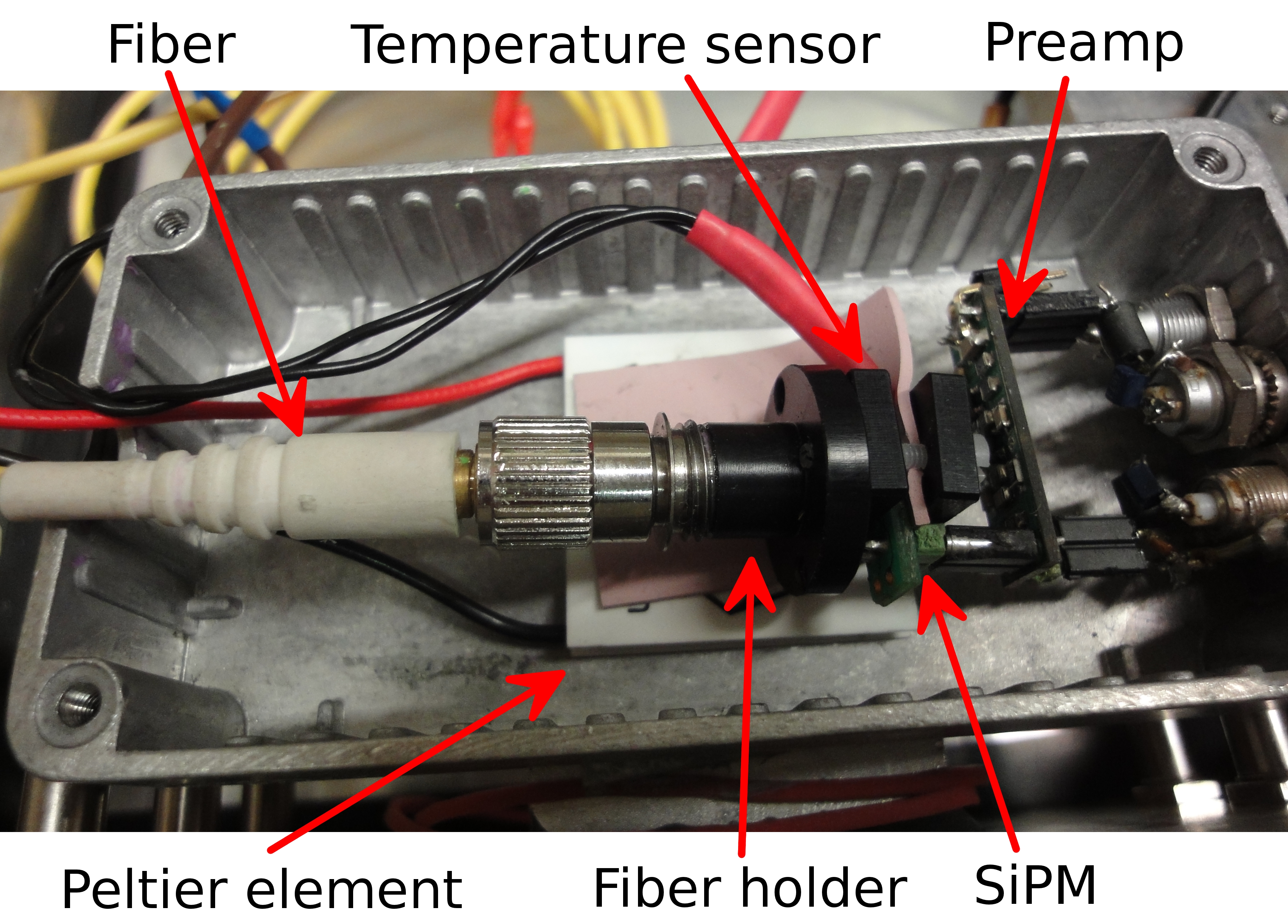}
  \includegraphics[width=.6\textwidth]{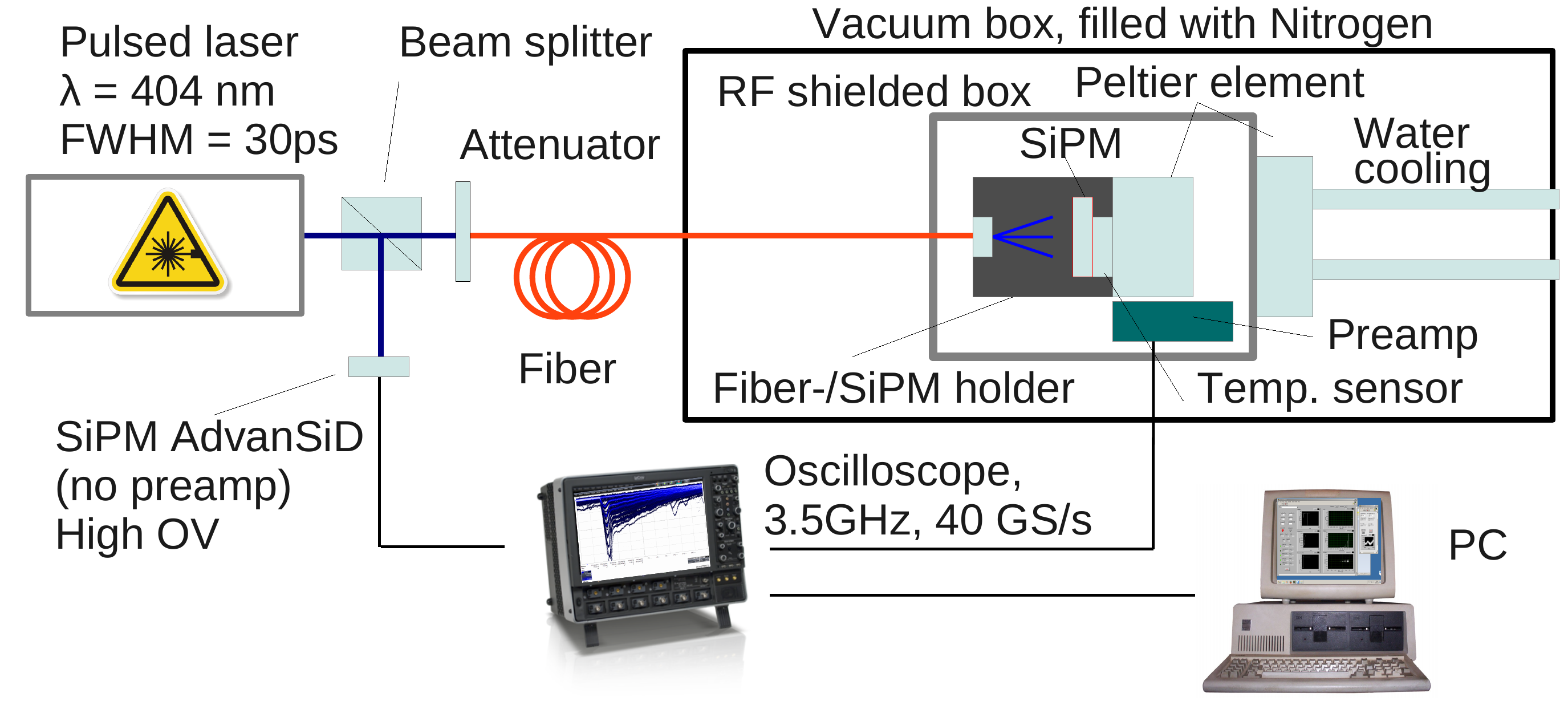}
  \caption{Picture of the RF-shielded box containing the tested SiPM and the preamplifier (left hand side). A schematic drawing of the whole setup (right hand side).}
  \label{fig:setup}
\end{figure}

In order to obtain the best detector time resolution for SciTil, the SiPM with the smallest intrinsic time spread needs to be found. Therefore, we performed a comparative timing performance study of SiPM from several vendors: AdvanSiD, Hamamatsu and Ketek. The figure of merit is the single photon time resolution (SPTR) of the devices. In order to obtain this value a semi-automatic setup was developed.
\paragraph{Setup} This setup is equipped with a picosecond laser from Advanced Laser Diode Systems (PIL040) with an emission wavelength of $\lambda$\,=\,404\,nm and a pulse width of $\simeq$\,30\,ps FWHM. The setup is shown in Fig.~\ref{fig:setup}. The laser beam is split into two paths. One path is used for providing a trigger signal and is led directly onto a 3SP50 SiPM from AdvanSiD. The SiPM is fully saturated for every laser pulse and its signal is read out directly without further amplification in order to obtain low time jitter. This trigger method was chosen, since better values of the SPTR were obtained as compared to using the trigger output of the laser itself.

The other path of the laser beam is directed onto the device to be measured. It passes a variable attenuator before it is coupled into a fiber, which is entering a vacuum box. The box was evacuated and filled with Nitrogen in order to avoid condensation of water on the electronic circuits. Inside the vacuum box, a single mode fiber was used to reduce the transit time spread of the light pulses. The fiber enters an RF-shielded box and is directed onto the SiPM. In order to obtain homogeneous illumination of the sensitive area, the fiber and the SiPM were mechanically coupled using a small plastic holder (see Fig.~\ref{fig:setup}). The SiPM were biased using a Keithley 6517A electrometer which was also used to measure the current in the SiPM. Additionally, a Keithley 617 electrometer was used to measure the bias voltage applied. Temperature stabilization was achieved by two Peltier elements which were coupled to a water chilled metal finger. The temperature was measured using a Pt100 temperature sensor, 
which was attached directly to the SiPM and was read out with a LakeShore 211 temperature monitor. An AMP-0611 preamplifier from Photonique was used to amplify the signal before it was fed into a LeCroy WavePro 735 Zi digital oscilloscope with a bandwidth of 3.5\,GHz and 40\,GS/s. A LabView program managed the measurements and controlled the oscilloscope for recording the data, the two electrometers, the Peltier elements and recorded the temperature.

\paragraph{Data taking} A measurement run was done at constant temperatures ($-10$\,$^\circ$C, 0\,$^\circ$C, 10\,$^\circ$C, 20\,$^\circ$C with an accuracy of\,$\pm$\,0.1\,$^\circ$C) varying the bias in small voltage steps. Starting from a bias voltage, slightly above the breakdown voltage, the program ramped the bias in steps of 0.1\,V or 0.2\,V respectively, dependent on the ratio of gain to bias voltage (e.g., for a small ratio, steps of 0.2\,V were applied). Several thousand events were recorded at each voltage step until a current limit (usually 8\,$\mathrm{\upmu A}$) was reached. Using the oscilloscope, the time difference relative to the trigger signal taken at 50\% of the amplitude, the amplitude, the area and the rise-/fall-time (10\%\,-\,90\%) of the waveform were recorded for each trigger. Additionally, the bias voltage, current and temperature for offline data analysis were measured and stored.

\paragraph{Data analysis} Automatic fitting of the pedestal and the single photon peak in the histograms of the signal amplitude was performed for all bias steps using a double Gaussian distribution. With this method, the single photon events are identified. Using the time stamps which correspond to single photon events (within a window of $\pm$\,2\,$\sigma$ of the single photon amplitude) the time delay spectrum of the signal relative to the trigger can be plotted. The resulting time stamp distribution was fitted with a Gaussian, giving a standard deviation which represents the SPTR.

\paragraph{Tested SiPM} SiPM of three different vendors, AdvanSiD, Hamamatsu and Ketek have been tested. A list of the devices and their parameters can be found in Table~\ref{tab:sipms}.

\begin{table}[tbp]
  \caption{List and detector parameters of the tested devices taken from the data sheets of the manufacturers. The PDE value indicated by $^\ast$ includes cross talk and after pulses.}
  \label{tab:sipms}
  \smallskip
  \centering
  \begin{tabular}{|lccccc|}
    \hline
    Manufacturer&AdvanSiD&Hamamatsu&\multicolumn{3}{c|}{Ketek}\\
    \multirow{2}{*}{Type}&SiPM3S&S10931&PM3375&PM3360&PM3350\\
    &P-50&-100P&-B72&-B66&-B63\\
    Total size [mm$^2$]&3$\times$3&3$\times$3&3$\times$3&3$\times$3&3$\times$3\\
    SPAD size [$\mu$m]&50&100&75&60&50\\
    Optical trenches&no&no&yes&no&yes\\
    Breakdown voltage [V]&$\simeq$\,35&$\simeq$\,70&$\simeq$\,23&$\simeq$\,23&$\simeq$\,23\\
    Darkcount rate [MHz]&$\leq$\,45&$\leq$\,12&$\leq$\,4.5&$\leq$\,4.5&$\leq$\,4.5\\
    Gain [$\times10^6$]&2.5&2.4&~14&~10&~6\\
    PDE at peak sensitivity [\%]&22&>\,70$^\ast$&>\,62&>\,60&>\,50\\
    Microcell capacitance [fF]&-&$\sim$2800&650&405&270\\
    \hline
  \end{tabular}
\end{table}

\subsection{Results}\label{sec:results}

\begin{figure}[tbp]
\centering
\includegraphics[width=.45\textwidth]{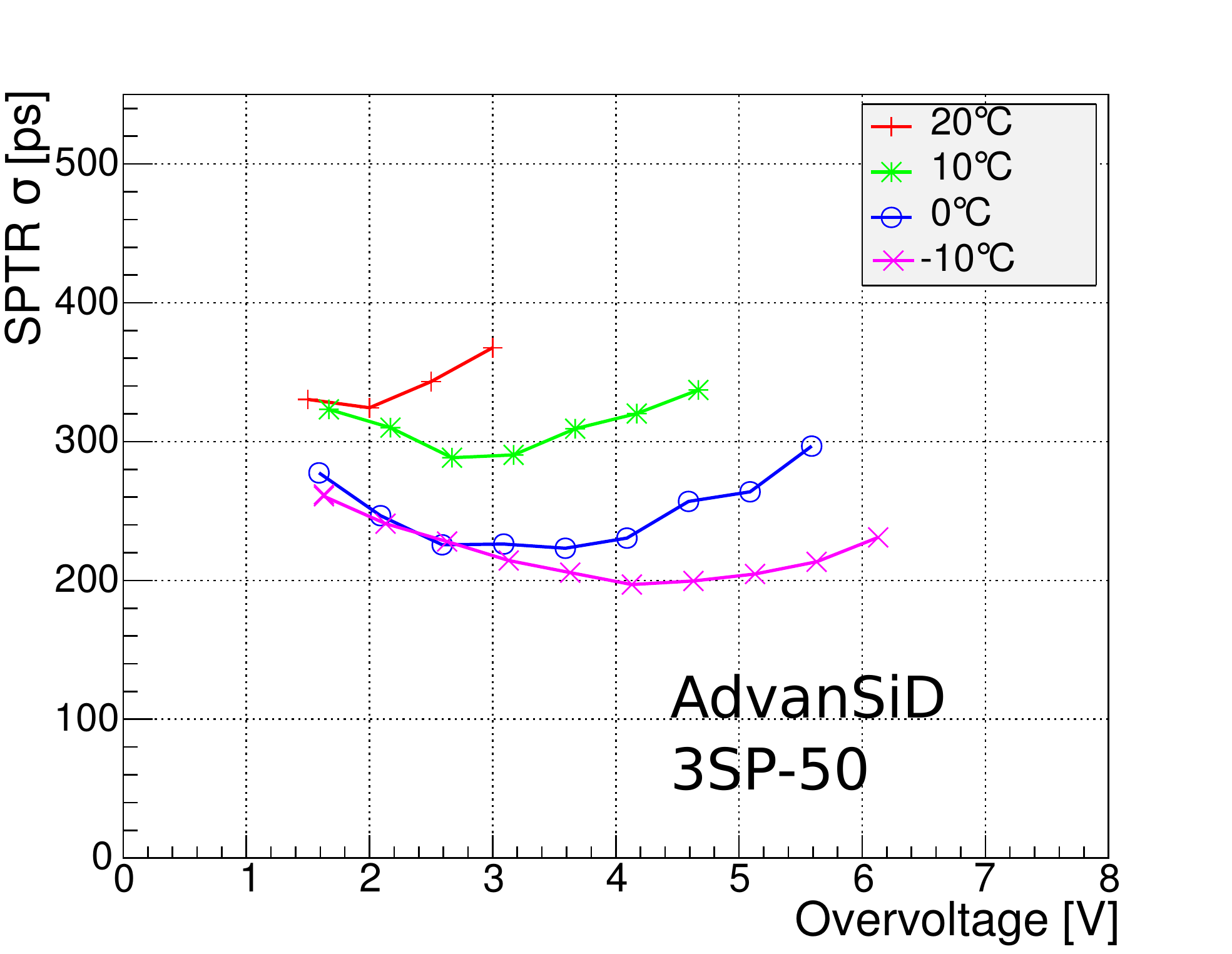}
\includegraphics[width=.45\textwidth]{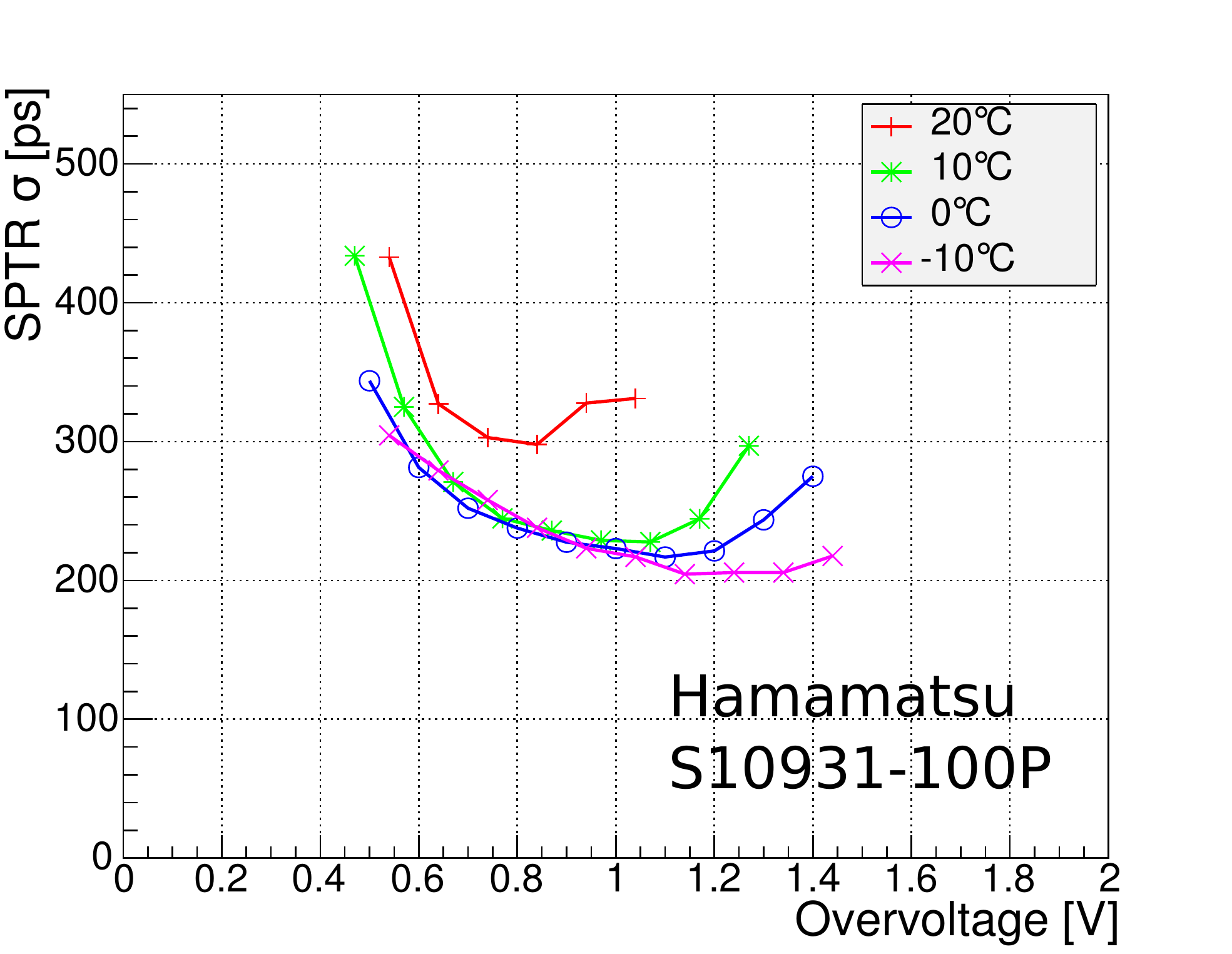}
\includegraphics[width=.45\textwidth]{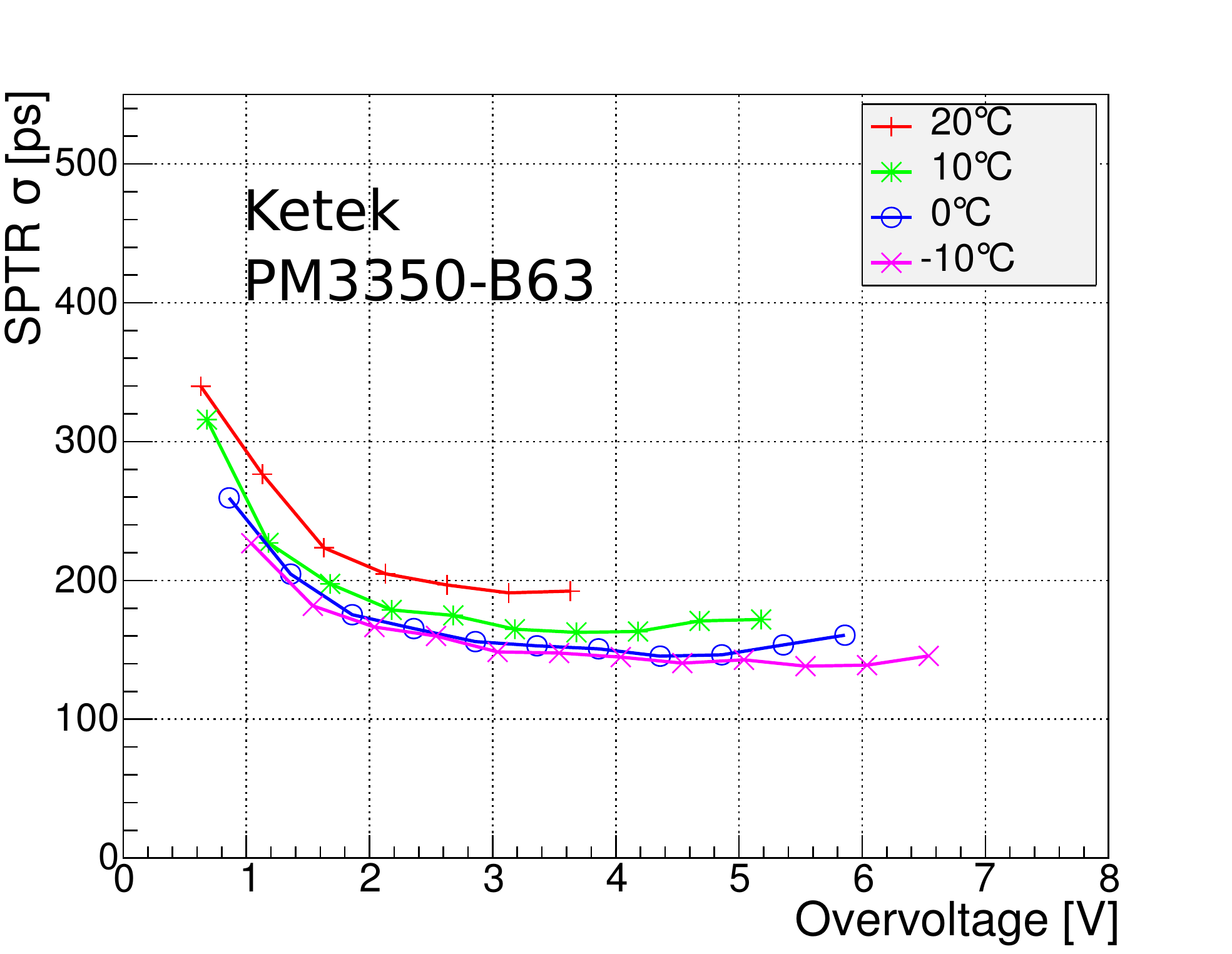}
\includegraphics[width=.45\textwidth]{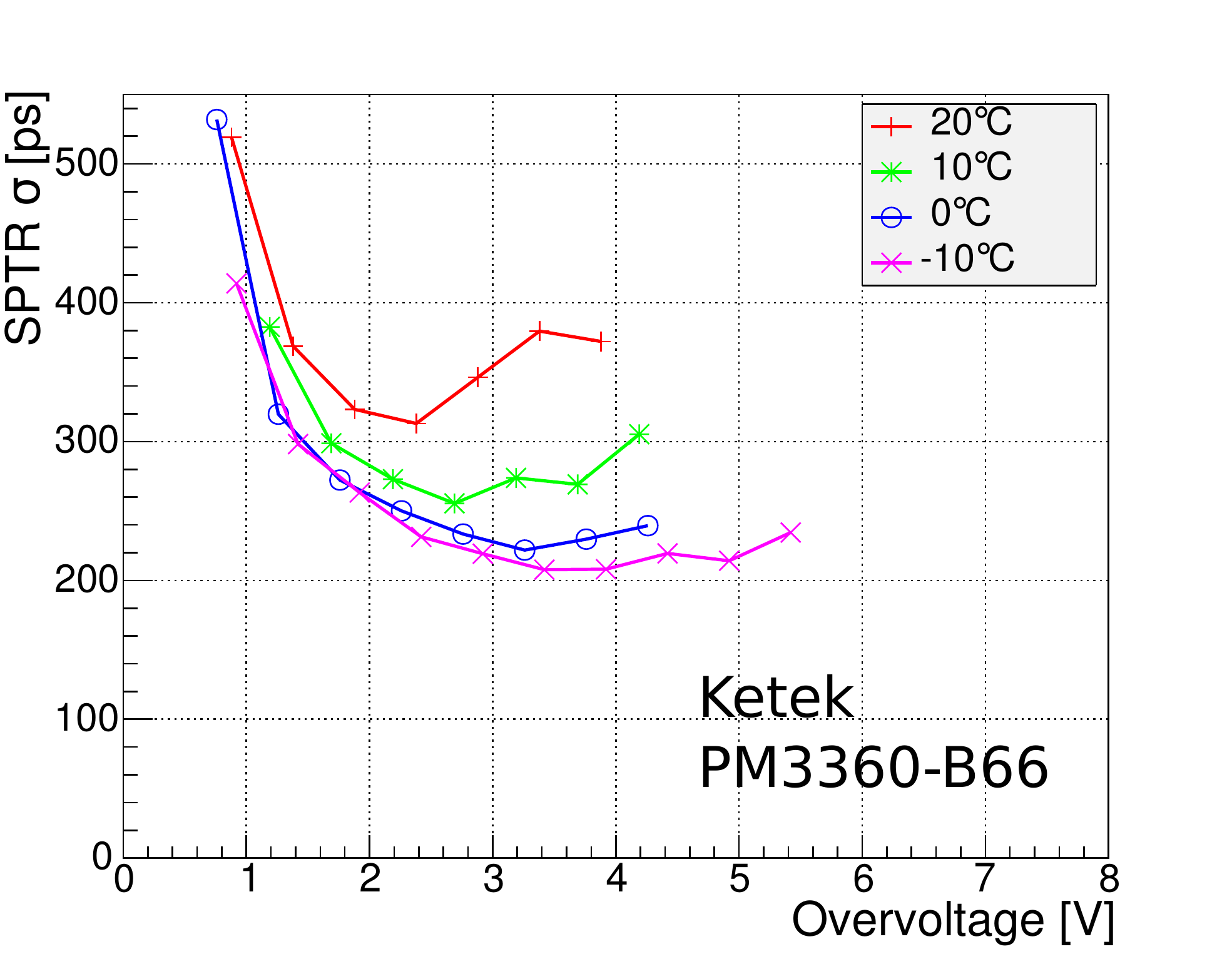}
\includegraphics[width=.45\textwidth]{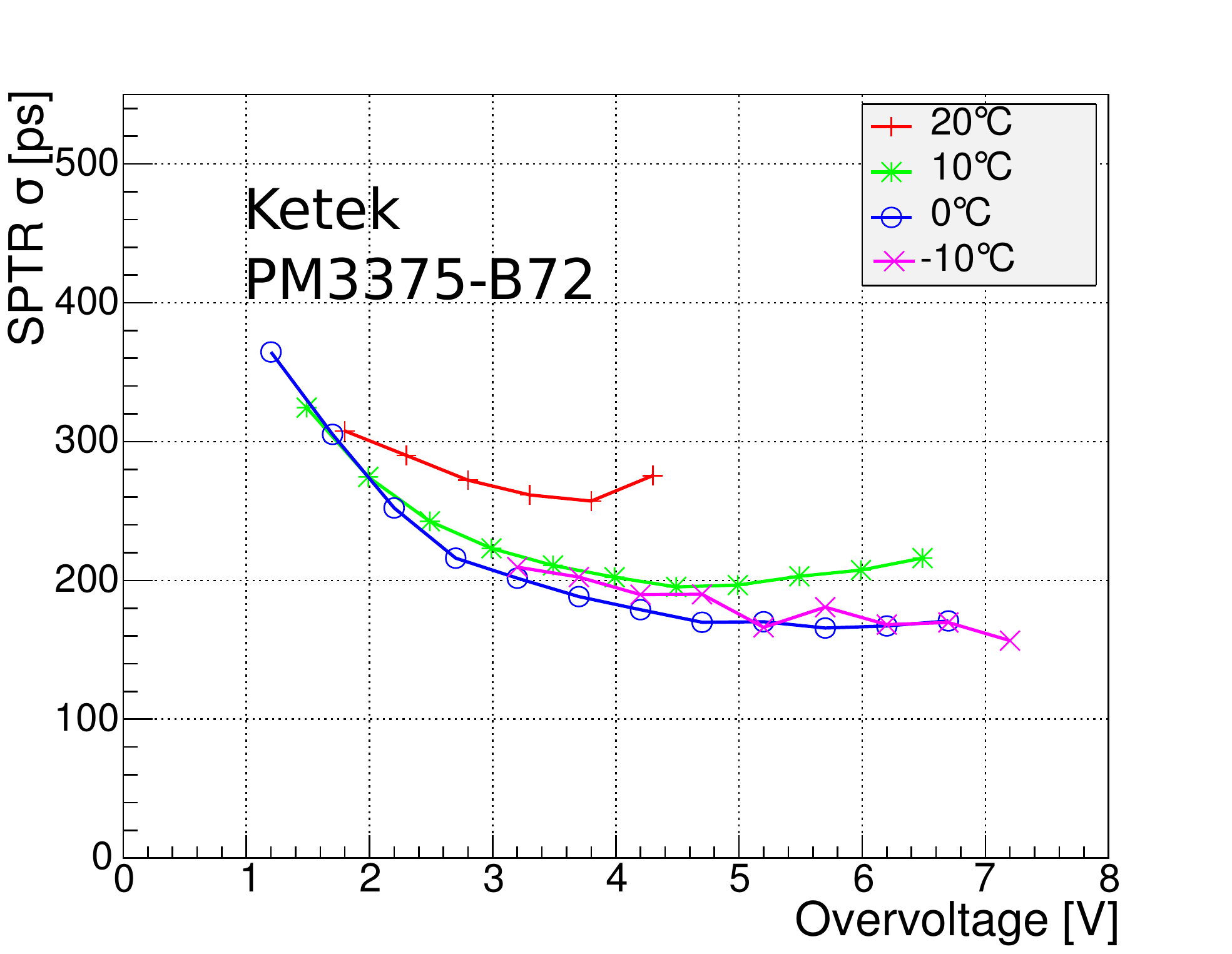}
\caption{Single photon time resolution of the tested SiPM. The lines are included to guide the eyes.}
\label{fig:sptr}
\end{figure}

The results of the SPTR measurements can be seen in Fig.~\ref{fig:sptr}, which shows the SPTR in dependency of the bias voltage and the temperature. The general trend is an improving time resolution with increasing bias voltage and with decreasing temperature, however, the improvement tends to saturate below 10\,$^\circ$C. The dependency on the bias voltage (which is proportional to the gain) is more prominent at higher temperatures, especially for the sensors from AdvanSiD and Hamamatsu and the Ketek PM3360. It indicates that the dark count rate, afterpulses and cross talk are factors influencing the time resolution of the SiPM as they are increasing with the bias voltage and decreasing with the temperature~\cite{Dinu2010}. This observation becomes more obvious when comparing the Ketek sensors PM3350 and PM3375 with optical trenches for reduction of cross talk with the PM3360 without optical trenches. Especially at 20\,$^\circ$C the difference of 
the SPTR-trend is significant.

It has to be noticed that the presented values of the SPTR do not represent absolute values but coincidence values of an asymmetric system corrupted by many factors in the intrinsic time resolution of the setup (e.\,g. trigger sensor operated at high temperature, the laser jitter, the preamplifier, the time precision of the first photon within a laser pulse is dependent on the laser pulse intensity, pick up noise) which becomes obvious when comparing the results with Ref.~\cite{Collazuol2007}. As the bias voltage range for the best SPTR at 20\,$^\circ$C is very small for the Hamamatsu sensors ($\sim$\,0.3\,V) compared to the Ketek sensors, the Ketek sensors would allow easier bias adjustment to obtain a good time resolution. This becomes important when considering the system time resolution of the several thousand channels planned for SciTil. Moreover, the Ketek sensors show the best timing performance over all bias settings and temperatures in this measurement. All three companies recently published new 
versions of their sensors, which will be tested in further measurements.

The SPTR is the parameter of interest when comparing the SiPM in terms of time resolution. Nevertheless, when using the SiPM in combination with the scintillator tiles, not only one but 100 photons per MIP are expected. Due to statistics the time resolution improves by a factor $1/\sqrt{\mathrm{N_{ph}}}$ with increasing number of detected photons N$\mathrm{_{ph}}$. Therefore, time resolutions well below 100\,ps should be achievable for all tested SiPM when detecting 100 photons.

\section{Conclusion and outlook}\label{sec:conclusion}
The time resolution of two square scintillators with different areas was measured using the Philips DPC. Best results were obtained when detecting photons centrally on the scintillator rims. For BC-408 a TOF resolution of 121\,ps (sigma) and for the EJ-228 a TOF resolution of 90\,ps (sigma) was obtained. Although the ratio of the sensitive area to the surface was smaller for EJ-228, a better TOF resolution was achieved. This indicates that the shorter rise- and decay-times of EJ-228 are significantly improving the time resolution of the scintillator. The time resolution of a single tile can be expected to be better than the measured TOF resolution. Further scintillator materials (EJ-200, EJ-204, EJ-232, BC-420) and geometries will be tested with our setup.

The time resolution of several SiPM was measured. Among the tested sensors the PM3350-B63 with optical trenches showed the best SPTR over a wide voltage range. Nevertheless, we will continue our tests with an improved setup and a broader variety of SiPM from more vendors.

As a next step, measurements with two analog SiPM attached to one scintillator tile will be performed. Furthermore, these systematic studies for optimizing the scintillator time resolution will be continued and finally, be tested at a particle accelerator facility.






\acknowledgments
We thank the Philips Technologie GmbH - Innovative Technologies acting through the team of Philips Digital Photon Counting for their support and interested participation in this research work. This work was partly supported by the EU Project HadronPhysics3 (project 283286).

\end{document}